\documentclass[paper=a4,fontsize=10pt,DIV=9]{scrartcl}

\usepackage{graphicx}
\usepackage{newtxtext,newtxmath,helvet}
\usepackage{cite,jabbrv}
\usepackage{caption}
\usepackage[T1]{fontenc}
\usepackage[tracking = true, letterspace = 16]{microtype}

\newcommand\authormark[1]{\textsuperscript{#1}}

\renewenvironment{abstract}%%
{\vskip1pc\noindent\textbf{Abstract:\space}}{}

\begin{document}

{\fontsize{16pt}{18.5pt}\sffamily\bfseries\selectfont\raggedright\textls{Optical modelling of accommodative light field display system and prediction of human eye responses}\par\vskip.15in}

{\raggedright\sffamily\bfseries\scshape\large {\boldmath}Yuta Miyanishi,\authormark{1,*} Erdem Sahin,\authormark{1} and Atanas Gotchev\authormark{1}\vskip1ex\par}

{\raggedright\small\itshape \authormark{1} Faculty of Information Technology and Communication Sciences, Tampere University, Korkeakoulunkatu 1, 33720 Tampere, Finland\par}

{\raggedright\footnotesize\itshape{}\authormark{*}yuta.miyanishi@tuni.fi\par}

{\raggedright\footnotesize\itshape{}https://research.tuni.fi/3dmedia/\par}

% \address{\authormark{1} Faculty of Information Technology and Communication Sciences, Tampere University, Korkeakoulunkatu 1, 33720 Tampere, Finland}

% \email{\authormark{*}yuta.miyanishi@tuni.fi} %% email address is required

% \homepage{https://research.tuni.fi/3dmedia/} %% author's URL, if desired


\begin{abstract}
The spatio-angular resolution of a light field (LF) display is a crucial factor for delivering adequate spatial image quality and eliciting an accommodation response. 
Previous studies have modelled retinal image formation with an LF display and evaluated whether accommodation would be evoked correctly. The models were mostly based on ray-tracing and a schematic eye model, which pose computational complexity and inaccurately represent the human eye population's behaviour. 
We propose an efficient wave-optics-based framework to model the human eye and a general LF display. 
With the model, we simulated the retinal point spread function (PSF) of a point rendered by an LF display at various depths to characterise the retinal image quality. 
Additionally, accommodation responses to rendered LF images were estimated by computing the visual Strehl ratio based on the optical transfer function (VSOTF) from the PSFs. 
We assumed an ideal LF display that had an infinite spatial resolution and was free from optical aberrations in the simulation. We tested images rendered at 0--4 dioptres of depths having angular resolutions of up to 4x4 viewpoints within a pupil. 
The simulation predicted small and constant accommodation errors, which contradict the findings of previous studies. 
An evaluation of the optical resolution of the rendered retinal image suggested a trade-off between the maximum resolution achievable and the depth range of a rendered image where in-focus resolution is kept high. 
The proposed framework can be used to evaluate the upper bound of the optical performance of an LF display for realistically aberrated eyes, which may help to find an optimal spatio-angular resolution required to render a high quality 3D scene. % (273 words)
\end{abstract}

%%%%%%%%%%%%%%%%%%%%%%%%%%  body  %%%%%%%%%%%%%%%%%%%%%%%%%%
\section{Introduction}
%ES:Explanation of VAC?: defocus blur addresses the optical (display) source plane, but the disparity addresses virtual image depth, is not this the main reason of VAC?
%ES: Distinguish  retinal blur vs. retinal defocus blur, not all "image blur" drives accommodation.
Accommodation is the mechanism of adjusting the refractive power of the crystalline lens in the eye. A blur pattern in a retinal image drives accommodation so that the retinal image is `sharp' or `best focused' about the fixated object. 
Accommodation is neurally coupled to binocular vergence, which is a function that rotates the two eyes in opposite directions simultaneously to fuse a visual target at a particular depth into a single image; therefore, an accommodation response is evoked even by a stimulus that elicits only a vergence response and vice versa~\cite{Bharadwaj2017,Inoue1997-qw,Torii2008-bp}. 
In a natural visual environment, this neural link supports quick and robust responses to visual targets so that a viewer can maintain a stable perception of 3D scenes. 
When observing a stereoscopic image rendered by a conventional 3D display, however, the two signals that drive accommodation may conflict. 
This is due to the blur pattern on the retina making the eye always focus at the optical distance to the visual target, including a conventional 3D display's surface, while vergence may widely vary depending on the rendered image depths and where the viewer fixates. 
This is known as the vergence-accommodation conflict, which causes visual discomfort and fatigue and may also hinder visual performance in some tasks~\cite{Hoffman2008-en,Johnson2016-qz}. 
Therefore, developing 3D displays that can provide focus information has become an important research topic. 

%Developing a 3D display that can provide focus information is a direct way to solve the conflict. Focus information refers to the depth information of an object in space that is estimated from variation of retinal blur depending on an accommodation state. 
%ES: is it vergence or disparity that cross-drives accommodation? It should be disparity?
%If a display can correctly replicate focus information of rendered 3D images, both disparity and retinal blur will drive accommodation to the rendered depth of the image and hense the vergence-accommodation conflict will be eliminated or at least mitigated. Furthermore, such display may create a defocus blur for not-in-focus objects as in the natural view so that visual perception may be improved. 

\begin{figure}
	\centering\includegraphics[width=133mm]{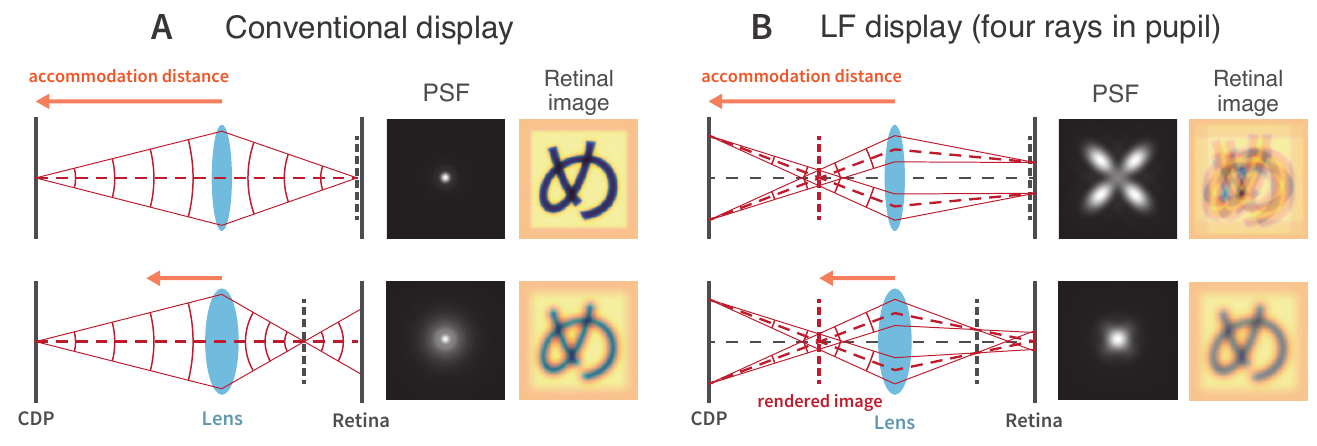}
	\caption{Basic mechanism of reproducing focus information by LF display. \textbf{A)} Conventional display and retinal images in a schematised eye. The retinal image is characterised by the point spread functions (PSFs) on the right of each diagram. Simulated retinal images of a letter are shown to the right of the corresponding PSFs. In the upper diagram, the eye accommodates at the display surface (indicated as the CDP; see Section~\ref{subsec:retinal_image_formation} for its definition), whereas the eye accommodates closer than the display surface in the lower diagram. \textbf{B)} An LF display that renders a point at a depth closer than the display with four rays in a pupil (only two are shown), PSFs, and simulated retinal images. In the upper diagram, the eye accommodates at the display surface. In the lower diagram, the eye accommodates at the rendered depth.}\label{fig:LFfocus}
\end{figure}

Light field (LF) displays modulate the position and direction of light rays, aiming to support all visual depth cues including binocular disparity, motion parallax, and focus information. 
This is attempted mainly by rendering a dense set of directional rays with a sufficiently high spatial and angular resolution. 
Fig.~\ref{fig:LFfocus} illustrates the basic mechanism of providing focus information by an LF display. 
When the viewer observes a conventional display (Fig.~\ref{fig:LFfocus}A), a focused and sharp retinal image is obtained only when the eye focuses on the display surface; otherwise the retinal image is blurred due to defocus. In other words, the focus information provided by the conventional display drives accommodation only at the display depth. 
An LF display, on the other hand, may provide more accurate focus information by rendering images with multiple rays that hit the eye's pupil, as shown in Fig.~\ref{fig:LFfocus}B. The rays' directions are controlled so that the rays intersect at the rendered point, imitating natural rays that would emanate from the point. When the eye focuses on the display surface, the individual images formed on the retina by the rays are optically focused; however, their superposition results in a separated or duplicated image that is expected to be recognised as a defocus blur by the visual system. On the other hand, when the eye focuses on the rendered point, the individual images by the rays are optically defocused but overlap perfectly. If the visual system recognises the latter as more `focused' than the former, the correct focus information about the rendered point is provided and accommodation should be driven to the rendered depth. 
However, the number of rays needed to drive accommodation correctly is not self-evident. 
% In the context of eye accommodation response, one can characterize LF displays in terms of number of different rays (viewpoints) hitting the pupil from a given object point.   
%

The idea of making accommodative LF displays was first proposed by Kajiki et al. in 1995 and validated by ray-tracing a simple optical model~\cite{Kajiki1996-ro}. 
Since then, several researchers have implemented the idea in various prototype displays and evaluated their performance in experiments, either by capturing a rendered image with a camera or measuring actual human viewers' accommodation responses~\cite{Miyanishi2019-ca,Honda2001-fq,Sudo2000-is,Nate2003,Mizushina2016-gk,Hiura2014-dy}. All these studies reported that accommodation responses were elicited by observing the rendered images, but the responses were not as close to the rendered depths as the responses to real objects. 
One reason for this may have been the low viewpoint density these prototypes achieved, which compromised three horizontally aligned viewpoints in the pupil at most. 

Alternatively to works experimenting with display prototypes, numerous theoretical studies have aimed at developing the analytical evaluation of factors affecting the optical performance of accommodative LF displays. 
%Even though recent progress in technology allows us to build prototypes that have higher viewpoint density, it is not realistic to test all implementation types of LF displays in combination with countless combinations of the display's optical setup. Moreover, display prototypes cannot be free from aberration and thus the effect of display aberration is inseparable from other factors in experimental results. 
%In contrast, theoretical evaluation allows analytical evaluation of how each factor affects the optical performance and delivered focus information, which may provide a crucial guide for designing accommodative LF displays. It also gives the supremum of an LF display's optical performance and accuracy of focus information by considering ideal conditions. 
% 
%\subsection{Theoretical studies about providing focus information by LF display}
% 
Huang and Hua proposed a framework to model an LF display and a human eye to simulate retinal images~\cite{Huang2017-bc}. 
They analytically described the formation of the retinal image and simulated it by ray-tracing a geometrical optical model that comprised the display part and the eye part. Optical resolution on the retina and the predicted accommodation responses were then evaluated by comparing simulated PSFs and modulation transfer functions (MTFs). 
They predicted accommodation responses considering an integral imaging display that creates 4--16 viewpoints in the pupil of a 3-mm diameter. The predicted responses appeared shifted towards the central depth plane (CDP: to be rigorously defined in Section~\ref{subsec:retinal_image_formation}) by approximately 0.1--0.2 dioptres. Higher viewpoint densities resulted in smaller predicted shifts. 
The model was improved further by taking into account the positional sampling and finite pixel resolution in the CDP~\cite{Huang2019-gd}, by generalising it for computational multi-layer LF displays~\cite{Xu2020-em}, and by unifying the framework to serve both integral imaging displays and computational multi-layer displays~\cite{Xu2020-rc}. 

Qin et al. proposed a model aimed at simulating retinal images rendered by a microlens-array-based near-eye LF display, taking into account off-axis ray propagation and using a rigorous calculation of diffraction~\cite{Qin2019-be}. 
Retinal images were simulated by ray-tracing a geometrical optical model that comprised pixels, microlenses, and the Arizona eye model; they later reported predicted accommodation responses to images rendered by that type of LF display~\cite{Qin2020-bi,Qin2020-db}. Their model had a fixed viewpoint density of 45 viewpoints over the 4-mm diameter pupil and varying CDP depth. 
%...by modifying the geometrical optical model. The accommodation responses were predicted by the Strehl ratio, which is a simple optical metric defined by the ratio of the peak height of an aberrated PSF to that of the corresponding diffraction-limited PSF. 
%They tested three CDP depths (0,2 and 4 diopters) and rendered image depths of 0--4 diopters. 
The predicted accommodation responses did not generally match the rendered image depths, and significant shifts of the predicted accommodation responses toward the CDP depth were observed. 
%in the simulation results especially for the CDP depths of 0D and 4D. The reported shifts were apparently much bigger than these reported in the study by Huang and Hua in 2017; for example, although the parameters Qin et al. and Huang and Hua tested do not match, Qin et al. reported shifts that were bigger than 0.5D for a rendered image closer than the CDP by 1D while Huang and Hua reported only 0.1--0.2D of the shifts for an image at the same depth relative to the CDP. 

%This dissimilarity between the two studies' results might be explained by the primary difference between the testing parameters of the two studies, namely the viewpoint density. However, Huang and Hua reported smaller shifts in the predicted accommodation response for viewpoint densities as high as 16 viewpoints in a 3-mm pupil; thus a very small shift is expected in a predicted accommodation response for even higher high viewpoint density of 45 viewpoints in a 4-mm pupil. 

%\subsection{Problem definition}
%ES: What is the "problem"? The below text is still more like a literature review. This part should be merged with the actual model/simulation we develop, which is now described in "Modelling and evaluation of LF display and human eye"
The methods reviewed above pose certain problems and issues that prevent their direct and systematic use for characterising LF displays and predicting the accommodation response to rendered images. First, these methods are based on ray-tracing, which is computationally inefficient and not suitable for generalisations. Second, the Arizona eye model has been commonly used as the required geometrical optical model of the human eye. This single model replicates chromatic effects and the `average' aberration of human eyes, but it does not represent the peculiarities of the population of human eyes. A handy customised eye model for this purpose is not available~\cite{Tabernero2017-qj}. 
Third, previous studies have not properly modelled and evaluated the accommodation responses for the case of polychromatic light, simulating instead either monochromatic retinal images or coarse spectral sampling. However, polychromatic effects are potentially important, since the human eye suffers from longitudinal chromatic aberration (LCA) greatly and the visual system makes use of it to drive accommodation~\cite{Cholewiak2018-we,Del_Aguila-Carrasco2020-oq}. 

%Another reason is that humans are always surrounded by polychromatic light in natural environments and the visual system must be tuned to polychromatically illuminated scenes in perceiving. Therefore, evaluation of a rendered image should not be done at a single wavelength but it should incorporate polychromatic light and chromatic effects. 
%To achieve the polychromatic evaluation, it is common to calculate a hypothetical polychromatic PSF as the superposition of monochromatic PSFs~\cite{Ravikumar2008-ni}. Monochromatic PSFs are often weighted by a luminosity function to reflect the visual effectiveness of each wavelength of light. 

% None of the previous studies properly conducted polychromatic evaluation. 
%None of the previous studies properly modelled and evaluated the polychromatic effects. 
%Qin et al. simulated retinal images only monochromatically~\cite{Qin2020-bi,Qin2020-db}. 
%The spectral sampling in Huang and Hua's simulation was relatively coarse, given that the PSFs and MTFs were calculated at five wavelengths from 470 nm to 650 nm~\cite{Huang2017-bc}. They even reduced the spectral sampling to three wavelengths in a recent study~\cite{Xu2020-em}. That may have made the polychromatic MTFs in the results less accurate, because a spectral sampling of 50 nm intervals may cause a large artefact on the PSF for a slightly aberrated eye~\cite{Ravikumar2008-ni}. 

The prediction of accommodation responses in previous studies may have also been less accurate because of the prediction measures used. These have been based either on the Strehl ratio or on values picked from through-focus MTFs at specific spatial frequencies; the latter does not give a single prediction of the accommodation error but different accommodation errors at different frequencies. 
On the other hand, the Strehl ratio, while a useful metric of optical performance for slightly aberrated optics, has demonstrated relatively poor performance as an accommodation response predictor~\cite{Cheng2004-ob}. 
%Instead, there are several optical metrics that are known to predict the empirical accommodation response well, and some of them are also good predictors of visual performances~\cite{Cheng2004-ob,Marsack2004-iq,Buehren2006-lj,Tarrant2010-wy}. In the current study, we used the visual Strehl ratio based on the optical transfer function (VSOTF). 

The primary purpose of the current study was to develop a novel framework to model an LF display and a human eye that does not require ray-tracing and can represent the population of aberrated human eyes. 
Secondarily, the study was aimed at simulating the retinal images rendered by an LF display so that (1) the accommodation responses to the images would be predicted and (2) optical resolution of the focused retinal images would be evaluated. 

\section{Modelling and evaluation of LF displays} 
LF displays are essentially aimed at reproducing the spatio-angular intensity distribution of light in a 3D scene. An LF propagating in the half space is usually represented by a four-dimensional function $L(s,t,u,v)$, where two parallel $(s,t)$ and $(u,v)$ planes parametrise the intensity variations along the spatial and angular dimensions of the LF, respectively~\cite{Balram2016-tg}. 
Different LF display technologies set these two planes differently. 
In multi-perspective projection-based LF displays, the two planes are associated with the plane where the projectors are located and the screen plane where the rays recombine~\cite{Bregovic2019-ec}. 
In integral imaging and super multiview displays, a light source plane and a direction-dependent light modulation plane form a pair of planes. 
In the current study, a microlens-array-based integral imaging display was modelled as a representative LF display technique, in line with other related studies, which have considered the same display principle~\cite{Huang2017-bc,Huang2019-gd,Xu2020-rc,Qin2019-be}. 
Other LF display techniques can be also investigated by matching their optical structure and light-modulation principle with the corresponding LF parametrisation~\cite{Huang2017-bc,Xu2020-rc}. 

\begin{figure}
    \centering
    \includegraphics[width=133mm]{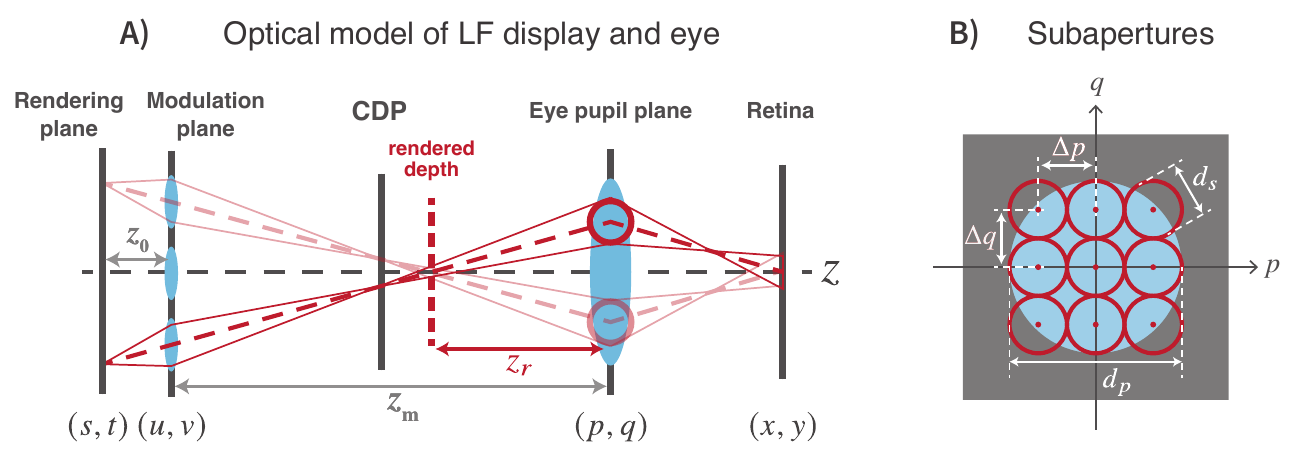}
    \caption{\textbf{A)}~An overview of the optical model of an LF display and a human eye. The model consists of six planes: the rendering plane, the modulation plane, the CDP (central depth plane), the depth plane of a rendered image, the eye pupil plane, and the retina. Two rays that render a point image to the right of the CDP are shown. The circles in the eye pupil plane indicate the subapertures formed by the two rays. \textbf{B)}~Example of subaperture arrangement for simulation. This example is for a viewpoint density of $3{\times}3$ viewpoints in the pupil. $d_p$ and $d_s$ refer to the diameters of the pupil and the subapertures, and ${\Delta}p$ and ${\Delta}q$ are the intervals of the subapertures in the $p$-- and $q$--directions.}
    \label{fig:modeloverview}
\end{figure}

\subsection{Retinal image formation}
\label{subsec:retinal_image_formation}

Fig.~\ref{fig:modeloverview} illustrates the retinal image formation process for an integral-imaging-based LF display. 
The process begins in the \emph{rendering plane} which represents a physical source of light. For multiview and integral imaging displays, the rendering plane corresponds to the LCD or OLED screen behind the microlenses. 
The \emph{modulation plane} in front of the rendering plane controls the directions of the emitted light with an optical structure such as a lenslet or a parallax barrier. Light emanating from a point source in the rendering plane forms a beam after being spatially limited and typically refracted in the modulation plane. 
The \emph{central depth plane} (CDP) is the optical conjugate of the rendering plane that is imaged by the modulation plane. 
% The optical distance from the eye to the CDP (CDP depth) is dependent on the optical setup of the rendering and modulation planes. If the modulation plane is implemented by microlenses, the CDP depth is determined from the microlenses' focal length and the distance between the two planes. If the modulation plane works as pinholes or a parallax barrier, the CDP is approximated to be at infinity. 
Light rays that emanate from the rendering plane and are modulated on the modulation plane intersect in a plane that includes the \emph{rendered 3D image} to render a 3D image point, imitating ideal rays that would emanate from that point. Therefore observers are expected to accommodate at this point (see Fig.~\ref{fig:LFfocus}). 
Each beam of light from a point source in the rendering plane then reaches the \emph{eye pupil plane}, which is a hypothetical plane that models the optical modulation in the eye, and it forms a \emph{subaperture}. 
% The amplitude distribution due to each beam on the eye pupil plane defines the \emph{subaperture}. 
The beam may be cropped by the pupil and is refracted, forming the image of the point source on the \emph{retina}, namely the \emph{elemental PSF}. The final retinal image of the rendered point is the incoherent superposition of the elemental PSFs formed by all rays that enter the eye through the pupil. The number of rays that hit the pupil, which is identically the number of subapertures, is the parameter of the greatest interest in evaluating an accommodative LF display. 

The position, shape, and size of the subapertures are determined by the setup of the rendering plane and the modulation plane. 
Let $z_r$ and $z_m$ be the depth of a rendered point and the depth of the modulation plane, respectively (Fig.~\ref{fig:modeloverview}A). 
% To render a point at a depth $z_r$ by the $n$-th beam that hits the pupil at $[p_n,q_n]$, the beam is parametrised using $(s,t)$ and $(u,v)$ with the depth of the modulation plane $z_m$, the distance between the modulation plane and the rendering plane $z_d$ and the depth of the rendered point $z_r$. Specifically, 
The positional interval of subapertures $[\Delta{}p,\Delta{}q]$ in Fig.~\ref{fig:modeloverview}B is geometrically linked to the sampling interval in the modulation plane $[\Delta{}u,\Delta{}v]$ by 
\begin{equation}
\begin{bmatrix}
{\Delta}p\\
{\Delta}q
\end{bmatrix}=\bigg|\frac{z_r}{z_r-z_m}\bigg|
\begin{bmatrix}
{\Delta}u\\
{\Delta}v
\end{bmatrix}\text{,}
\end{equation}
assuming infinitely high resolution on the rendering plane ($(s,t)$ plane), which is equivalent to imposing no positional constraint in determining the locations of the beam emitters. 
The shape and size of each subaperture is dependent on the modulation structure of the modulation plane. 
Assuming the origin of the pupil plane is at the centre of the pupil, the amplitude function on the $n$-th subaperture $A_\mathrm{elem}(p,q;n)$ is derived as  
\begin{equation}\label{eq:Aelem}
    A_\mathrm{elem}(p,q;n) = \Big|\,\mathfrak{Fr}_{z_m}\big[P_m(u,v)\,\mathfrak{Fr}_{z_0}[\delta(s-s_n,t-t_n)]\big]\Big|\text{,}
\end{equation}
where $\mathfrak{Fr}_z[U(\xi,\eta)]$ denotes the Fresnel diffraction integral~\cite{goodman2017} as a convolution between an input field $U(\xi,\eta)$ and the convolution kernel for propagation distance $z$, $P_m(u,v)$ represents the modulation function of the modulation structure for the corresponding beam, $\delta(s-s_n,t-t_n)$ denotes the Dirac delta function at $[s_n,t_n]$ modelling the point source in the rendering plane, and $z_0$ and $z_m$ refer to the distance from the rendering plane to the modulation plane and the distance from the modulation plane to the eye pupil plane. 
The modulation function $P_m(u,v)$ is a simple aperture function or a slit function for displays that modulate light with pinholes or a parallax barrier. For an integral imaging display that controls light's directions by lenslets, the modulation function includes the phase modulation term corresponding to a microlens, which is a chirp function.

In the current study, a hypothetically ideal modulation in the modulation plane was assumed. That is to say, the effects of diffraction and aberration due to the optical modulation structure were neglected, and a point source in the rendering plane was assumed to be imaged in the CDP as an ideal point so that the light sources on the rendering plane can equivalently be treated as ideal point sources in the CDP. 
This assumption separates the effects of a display's optical structure from the modelled retinal image formation process. Such a model gives ideal simulation results, which however apply to all LF displays that share the same critical parameters such as ray density and CDP depth, and provide the theoretical maximum of the LF displays' performance. 
In our simulations, a subaperture $A_\mathrm{elem}$ was geometrically determined instead of being derived rigorously from Eq.~\ref{eq:Aelem} as shown schematically in Fig.~\ref{fig:modeloverview}A. 
We first defined three viewpoint densities for the simulation, namely 2$\times$2, 3$\times$3, and 4$\times$4 viewpoints in a 3-mm diameter pupil. To simulate retinal images for a viewpoint density selected from these, the size and positions of the subapertures were determined from the viewpoint density as illustrated in Fig.~\ref{fig:modeloverview}B. Specifically, the subapertures were set to circular functions that were tangential to each other on a rectangular grid in the eye pupil plane filling the pupil; thus, the diameter of the subapertures $d_s$ was equal to the intervals of the subapertures ${\Delta}p$ and ${\Delta}q$. 

\begin{figure}[t]
	\centering\includegraphics[width=133mm]{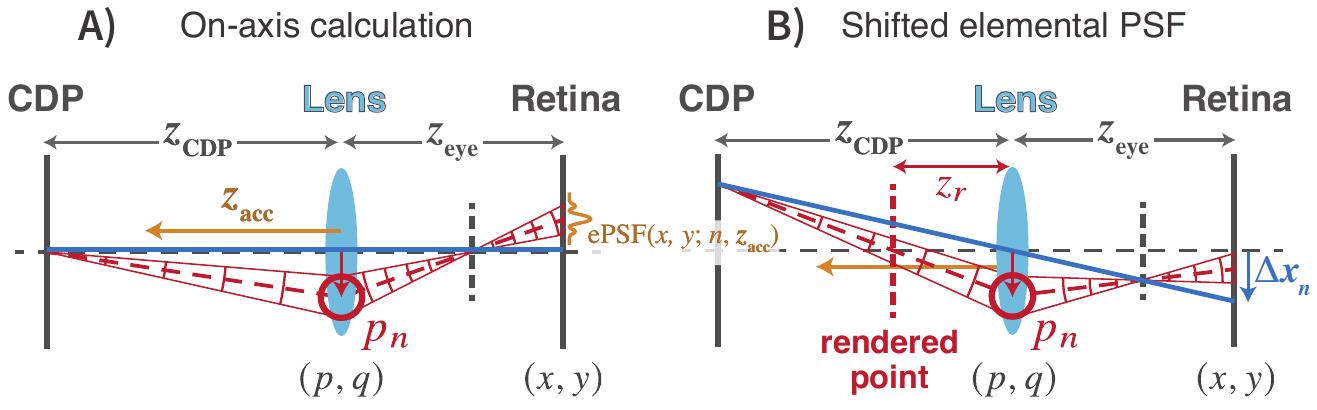}
	\caption{Lateral shift of an elemental PSF. The diagrams show only the $x$ and $p$ dimension for simplicity. The red circle in the lens denotes the $n$-th subaperture and $p_n$ is the position of its centre. \textbf{A)} The elemental PSF ($\mathrm{ePSF}(x,y;n,z_\mathrm{acc})$) as the image of an on-axis point at the CDP. \textbf{B)} The shifted elemental PSF when the LF display renders a point at the depth $z_r$. The shift $\Delta{}x_n$ is geometrically determined by $z_\mathrm{CDP}$, $z_\mathrm{eye}$, $z_r$, and $p_n$. }\label{fig:elemPSFshift}
\end{figure}

Under the assumption of a point source at the CDP and the paraxial approximation, an elemental PSF due to a given beam is obtained by Fourier-transforming the corresponding generalised pupil function, which is nonzero within the corresponding subaperture, as illustrated in Fig. \ref{fig:elemPSFshift}A~\cite{goodman2017}. 
Specifically, the elemental PSF for the $n$-th subaperture at a wavelength $\lambda$ for an eye with a nominal accommodation distance $z_\mathrm{acc}$, namely $\mathrm{ePSF}(x,y;\lambda,n,z_\mathrm{acc})$, is described as 
\nopagebreak
\begin{equation}
\mathrm{ePSF}(x,y;\lambda,n,z_\mathrm{acc}) \propto \Big|\left.\mathcal{F}\!\left[P_\mathrm{sub}(p,q;\lambda,n,z_\mathrm{acc})\right]\right|_{\big(\frac{x}{\lambda{}z_\mathrm{eye}},\frac{y}{\lambda{}z_\mathrm{eye}}\big)}\Big|^2
\end{equation}
where $\mathcal{F}[\cdot]$ denotes the Fourier transformation, $P_\mathrm{sub}(p,q;\lambda,n,z_\mathrm{acc})$ is the generalised pupil function of the $n$-th subaperture defined over the whole eye pupil, and $z_\mathrm{eye}$ is the optical distance from the eye pupil plane and the retina. The generalised pupil function is defined as 
\nopagebreak
\begin{equation}\label{eq:PupilFunc}
P_\mathrm{sub}(p,q;\lambda,n,z_\mathrm{acc}) = A_\mathrm{sub}(p,q;\lambda,n)\exp{}\Big[\frac{i2\pi{}}{\lambda}\mathit{W}(p,q;\lambda,z_\mathrm{acc})\Big]\text{,}
\end{equation}
\nopagebreak
where $A_\mathrm{sub}(p,q;\lambda,n)$ is the aperture function that describes the amplitude modulation and $\mathit{W}(p,q;\lambda,z_\mathrm{acc})$ is the wavefront aberration function that represents the phase modulation. The generalised pupil function, the aperture function, and the wavefront aberration function are all defined not only in the subaperture but in the whole eye pupil plane. 

The aperture function $A_\mathrm{sub}(p,q;\lambda,n)$ incorporates the amplitude variation of the incident light beam within the corresponding subaperture $A_\mathrm{elem}(p,q;n)$ and a Gaussian apodisation filter in the whole eye pupil plane $A_\mathrm{SCE}(p,q;\lambda) =10^{-\rho_{\lambda}(p^2+q^2)}$, which models the effect that light passing near the centre of the pupil stimulates cones much more strongly than light passing near the edge of the pupil, referred to as the Stiles--Crawford effect~\cite{Stiles1933-uu,Westheimer2008-rf}. The peak of the effectiveness of the entering light is assumed to be at the centre of the pupil (namely the origin in the $p,q$-plane), and the wavelength-dependent parameter $\rho_{\lambda}$ denotes the peakedness of the effect~\cite{Berendschot2001-it}. 
The resulting amplitude modulation can then be written as 
\begin{equation}
A_\mathrm{sub}(p,q;\lambda,n) = A_\mathrm{pupil}(p,q)A_\mathrm{SCE}(p,q;\lambda)A_\mathrm{elem}(p,q;n),
\end{equation}
where $A_\mathrm{pupil}(p,q)$ represents the binary circular pupil function $A_\mathrm{pupil}(p,q)=\mathrm{circ}\Big(\frac{\sqrt{p^2+q^2}}{d_p}\Big)$ for a pupil with diameter $d_p$. 
% In the current study, we assumed the hypothetically ideal modulation in the modulation plane and an ideal point source in the CDP. Therefore the subaperture $A_\mathrm{elem}$ was geometrically determined as shown schematically in Figure~\ref{fig:modeloverview}A instead of deriving rigorously from Equation~\ref{eq:Aelem}. 
% We first defined three viewpoint densities for the simulation, namely 2$\times$2, 3$\times$3, and 4$\times$4 viewpoints in a 3-mm diameter pupil. To simulate retinal images for a viewpoint density selected from these, the size and positions of the subapertures were determined from the viewpoint density as illustrated in Figure~\ref{fig:modeloverview}B. Specifically, the subapertures were set to circular functions that were tangent to each other on a rectangular grid in the eye pupil plane filling the pupil; thus the diameter of the subapertures $d_s$ was equal to the intervals of the subapertures ${\Delta}p$ and ${\Delta}q$. 
%
%ES_v2: the distinction between subapert size and interval / angular sampling may be elaborated here as an importan factor that may affect the resulting PSF: then we say we consider the case where subapert size = subapert interval
The wavefront aberration function $W(p,q;\lambda,z_\mathrm{acc})$ incorporates all possible wavefront aberrations compared to the diffraction-limited imaging condition, including not only the fixed aberrations of the eye but also accommodation-dependent defocus and spherical aberration terms. The composition of the wavefront aberration function is further elaborated in Section~\ref{subsubsec:wavefront_aberrations}. 

The polychromatic PSF $\mathrm{ePSF_{poly}}(x,y;n,z_\mathrm{acc})$ for each subaperture is found by superposing the elemental PSFs at different wavelengths as

\nopagebreak
\begin{equation}
\mathrm{ePSF_{poly}}(x,y;n,z_\mathrm{acc}) = {\int}\mathrm{ePSF}(x,y;\lambda,n,z_\mathrm{acc})V(\lambda)\,d\lambda. 
\end{equation}
The monochromatic PSFs are weighted by a luminosity function $V(\lambda)$ to reflect the spectral visual effectiveness~\cite{ISO11664}. In the current study, the CIE physiologically relevant 2-degree luminosity function for photopic vision was used as $V(\lambda)$~\cite{Stockman2008-av}, principally because only foveal vision was under consideration. 
An elemental PSF calculated in this way represents the PSF of the on-axis point source (see Fig.~\ref{fig:elemPSFshift}A). To obtain the retinal PSF of a rendered point, the elemental PSFs are laterally shifted on the retina reflecting the rendered image depth $z_r$, the CDP depth $z_\mathrm{CDP}$, and $z_\mathrm{eye}$ (see Fig.~\ref{fig:elemPSFshift}B). The amount of shift ${\Delta}\mathbf{s}_n$ is determined as  
%ESv2: check refs above
\nopagebreak
\begin{equation}
{\Delta}\mathbf{s}_n = 
[\Delta{}x_{n},\Delta{}y_{n}]
= z_\mathrm{eye}\bigg(\frac{1}{z_r}-\frac{1}{z_\mathrm{CDP}}\bigg)
[p_n,q_n],
\end{equation}
where $[p_n,q_n]$ is the position of the $n$-th subaperture's centre. 

Finally, laterally shifted polychromatic elemental PSFs are superposed and the retinal PSF of the rendered point under a nominal accommodation distance $z_\mathrm{acc}$ is obtained as
\nopagebreak
\begin{equation}
\mathrm{PSF_{poly}}(x,y;z_\mathrm{acc}) = \sum_n\mathrm{ePSF_{poly}}(x-{\Delta}x_{n},y-{\Delta}y_{n};n,z_\mathrm{acc})\text{.}
\end{equation}

\subsubsection{Eye wavefront aberrations}
\label{subsubsec:wavefront_aberrations}

The wavefront aberration function $W(p,q;\lambda,z_\mathrm{acc})$ in Eq.~\ref{eq:PupilFunc} includes statistics-based monochromatic aberrations of an unaccommodated eye, the LCA as a wavelength-dependent defocus, the accommodation-related defocus, and the accommodation-dependent additional spherical aberration. 

A wavefront aberration function $W(p,q)$ can be described by a Zernike expansion that is a weighted sum of Zernike polynomials as 
\nopagebreak
\begin{equation}
    W(p,q)=\sum_{m,n}c^m_nZ^m_n(p,q)\text{\ ,}
    \label{eq:wavefront_aber_zernike}
\end{equation}
where $Z^m_n(p,q)$ is a Zernike polynomial, $c^m_n$ is the corresponding Zernike coefficient, and $m$ and $n$ ($n,m{\in}\mathbb{Z}; n{\geq}|m|{\geq}0$) denote the highest order of the polynomial's radial component and the azimuthal frequency of the polynomial's sinusoidal component, respectively~\cite{Thibos2000-qp}. 
With a Zernike expansion, any wavefront aberration function is represented by a set of the Zernike coefficients. 
%For the sake of simplicity, hereafter we omit the indexes $m$ and $n$. For an aberrated eye, each coefficient $c$ is a sum of four elements ($c_\mathrm{accDF}$, $c_\mathrm{accSA}$, $c_\mathrm{aber}$, and $c_\mathrm{LCA}$) which represent the above-mentioned accommodation-related defocus, spherical aberration, and monochromatic and chromatic aberrations, respectively. 

A wavefront aberration function includes \emph{defocus} of the optical system, which is zero if the system satisfies the imaging condition. In other words, a non-zero defocus means a signed longitudinal shift of the nominal focal point from the CDP depth. 
Given the nominal accommodation distance $z_\mathrm{acc}$ is provided in metres, the defocus in dioptres $D_\mathrm{acc}(z_\mathrm{acc})$ is found by the following equation: 
\begin{equation}
D_\mathrm{acc}(z_\mathrm{acc})=\frac{1}{z_\mathrm{acc}}-\frac{1}{z_\mathrm{CDP}}\text{,}\label{eq:Dacc}
\end{equation}
where $z_\mathrm{CDP}$ is the CDP depth in metres. 
% A dioptric defocus (often also called as `spherical equivalent') $S$ is converted to the coefficient $c^0_2$ of the Zernike defocus term $Z^0_2$ by the formula: 
A dioptric defocus $D_\mathrm{acc}(z_\mathrm{acc})$ contributes to the Zernike defocus term $\textstyle Z^0_2(p,q)$ of the wavefront aberration function. The corresponding coefficient is found through the following equation: 
\begin{equation}\label{eq:diopter2um}
c_\mathrm{acc}(z_\mathrm{acc})=\frac{{d_p}^2}{16\sqrt{3}}D_\mathrm{acc}(z_\mathrm{acc})\text{\ ,}
\end{equation}
where $d_p$ is pupil diameter in millimetres and $S$ is defocus in dioptres~\cite{Thibos2004-vg,ANSI-Z80-28}. Note that $c_\mathrm{acc}(z_\mathrm{acc})$ is in micrometres. 
%The coefficient vector $c_\mathrm{accDF}$$ has the value $c_\mathrm{acc}(z_\mathrm{acc})$ for the component that corresponds to the Zernike defocus term and zero for all other components. 

Accommodation primarily changes the shape of the crystalline lens, resulting in an increase in its refractive power and also a systematic decrease in the primary spherical aberration while the other types of aberration statistically show little change with accommodation~\cite{He2000-eq,Cheng2004-dn,Lopez-Gil2008-qg,Del_Aguila-Carrasco2020-oq}. 
Cheng et al. measured changes of aberrations with accommodation and reported that the change in spherical aberration was proportional to the change in accommodation defined by the Zernike defocus~\cite{Cheng2004-dn}. 
Specifically, the reported slope of the change in the Zernike coefficient for primary spherical aberration $c^0_4$ was $-0.0435$ um/D for 5-mm pupils. 
The coefficient $c_\mathrm{accSA}$ reflects the contribution of such accommodation-dependent factor to the spherical aberration term $Z^0_4(p,q)$. For a pupil of 5-mm diameter, it is given as 
\begin{equation}
c_\mathrm{accSA} = \frac{-0.0435}{z_\mathrm{acc}}\text{.}    
\end{equation}

%The term $c_\mathrm{aber}$ represents monochromatic aberrations of the eye at the unaccommodated state. 
Human eyes are considerably aberrated, and it greatly affects the quality of the retinal image. Therefore, modelling the human eye with simple ideal optics must over-estimate the retinal resolution. 
Furthermore, idealising the human eye in the model may also cause an incorrect prediction of the accommodation responses. This is because the empirical accommodation response as a function of the stimulus's optical distance is widely known to be S-shaped, while the response function of the idealised (diffraction-limited) eye model should be simply the identical line~\cite{Buehren2006-lj,Del_Aguila-Carrasco2020-oq,Labhishetty2021-zn}. This phenomenon, namely the \emph{lead and lag of accommodation}, is considered to be due to ocular aberrations and pupil constriction~\cite{Lopez-Gil2010-hf,Thibos2013-mf,Lopez-Gil2013-ap}.
Similarly, simulating a retinal image for one eye that has the average aberration of the human eye population may also give misleading results. This is because aberrations of the human eye population tend to distribute around zero and thus the hypothetical `average' eye is nearly free of aberrations~\cite{Thibos2002-jr}. In other words, the `average' eye does not represent the human eye population. 

\begin{figure}
    \centering
    \includegraphics[width=67mm]{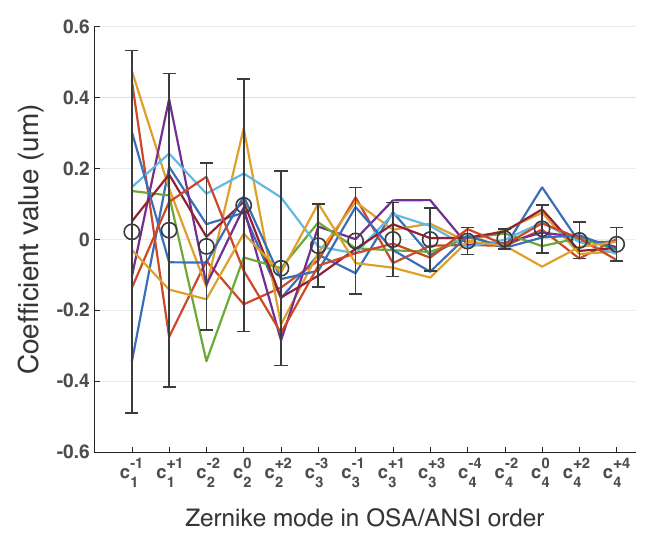}
    \caption{Zernike coefficient values for modes 1--13 in OSA/ANSI order. Circles and error bars indicate mean and $\pm{}2$ standard deviation of each coefficient value in the statistical model. Colored lines represent generated Zernike coefficient sets for each virtual eye.}
    \label{fig:EyesZernike}
\end{figure}

% For this reason, one has to evaluate multiple instances of the eyes the aberrations of which statistically follow the population of the human eyes in order to obtain simulation results that correctly demonstrate the responses of the eyes. 
To address this issue, we generated multiple instances of an aberrated eye that follow the statistics of measured human eye aberration. Simulated retinal images in these instances, which truly represent the human eye population, were evaluated. 
To generate corresponding sets of aberration coefficients, we used the statistical model of the aberration of the healthy human eye population by Thibos et al.~\cite{Thibos2002-en}. 
In the model, the coefficients for individual eyes are represented as multivariate Gaussian variables with the measured mean and variance of each coefficient and the covariances between all possible pairs of the coefficients. 
In this study, coefficients representing ten virtual aberrated eyes were generated from the given vector of the mean values of the coefficients and the variance-covariance matrix~\cite{Thibos2009-ba}. 
Fig.~\ref{fig:EyesZernike} shows the resulting Zernike coefficients of the generated virtual eyes and the mean and $\pm{}2$ standard deviation of each coefficient described in the statistical model. 
The mean coefficient values are close to zero for most of the Zernike modes, while the coefficient values are much larger for the generated eyes in absolute value. 

The chromatic aberrations, especially the LCA, also significantly affects the retinal image. The LCA causes the wavelength-dependent defocus of light --- specifically, shorter wavelengths of light images at points closer to the lens and vice versa. The defocus due to the LCA is always present in natural viewing, and the visual system utilises it to drive accommodation~\cite{Cholewiak2018-we,Del_Aguila-Carrasco2020-oq}. 
In contrast to monochromatic aberrations in the human eye population, the profile of the LCA is mostly common across individuals~\cite{Thibos1992-fx,Nakajima2015-fh}. The LCA is measured in dioptres and modelled at a wavelength $\lambda$ by 
\begin{equation}
    D(\lambda) = A\Bigg(\frac{1}{\lambda-c}-\frac{1}{\lambda_\mathrm{ref}-c}\Bigg)\text{,}
\end{equation}\nopagebreak
where $A = 633.26$ and $c = 214.10$~\cite{Thibos1992-fx}. The parameter $\lambda_\mathrm{ref}$ is a reference wavelength in nanometres, at which the retinal image is assumed to be nominally focused. The contribution of the LCA to the defocus term $\textstyle Z^0_2(p,q)$ can be found by converting $D(\lambda)$ to micrometres using Eq.~\ref{eq:diopter2um}.
%Since $D(\lambda)$ is in diopters, it must be converted into the coefficient of the Zernike defocus term $c_\mathrm{LCA}\!(\lambda)$ with Eq.~\ref{eq:diopter2um}. 
%The coefficients $c_\mathrm{LCA}$ are zero except the one corresponding to the defocus term $\textstyle Z^0_2$ which has the value $c_\mathrm{LCA}\!(\lambda)$. 

\subsection{Prediction of accommodation response}

The metrics for the prediction of accommodation responses incorporate not only the retinal image itself but also the subsequent neural factors, reflecting the fact that a retinal defocus blur is detected, processed, and analysed in the visual processing pathway from the retina to the cortex to decide the accommodation response. The visual Strehl ratio is an objective metric that involves the neural transfer function and is known to be one of the best reliable predictors of subjective focus and visual acuity~\cite{Thibos2004-vg,Cheng2004-ob,Marsack2004-iq,Tarrant2010-wy,Martin2011-ns,Ravikumar2012-rr}. 
While the Strehl ratio is defined as the ratio of a PSF's maximum to that of the diffraction-limited PSF, the visual Strehl ratio is computed in the frequency domain (namely the optical transfer function or the modulation transfer function), where it is weighted by a neural factor. 
In our setting, the visual Strehl ratio computed from the optical transfer function (VSOTF) is given by 
\begin{equation}
\mathrm{VSOTF} = \frac{\int_{-\infty}^{\infty}\int_{-\infty}^{\infty}\mathrm{NCSF}(f_x,f_y)\mathrm{OTF_{poly}}(f_x,f_y)df_xdf_y}{\int_{-\infty}^{\infty}\int_{-\infty}^{\infty}\mathrm{NCSF}(f_x,f_y)\mathrm{OTF_{DL}}(f_x,f_y)df_xdf_y}\text{,}
\label{eq:VSOTF}
\end{equation}%
where $\mathrm{OTF_{poly}}$ is the given OTF, $\mathrm{OTF_{DL}}$ is the OTF of the diffraction-limited optics, and $\mathrm{NCSF}$ is the neural contrast sensitivity function~\cite{Thibos2004-vg}.  
Specifically,  
\begin{equation}
\mathrm{OTF_{poly}}(f_x,f_y;z_\mathrm{acc})=\mathcal{F}\big[\mathrm{PSF_{poly}}(x,y;z_\mathrm{acc})\big]\text{,}
\end{equation}
and  
\begin{equation}
\mathrm{OTF_{DL}}(f_x,f_y) = \mathcal{F}\big[\mathrm{PSF_{polyDL}}(x,y)\big]\text{,}
\end{equation}
% where $\mathrm{PSF_{polyDL}}(x,y)$ is defined as the retinal PSF obtained by applying $z_\mathrm{acc} = z_\mathrm{CDP}$ and the terms $c_\mathrm{aber}$ are set to zero while finding the wavefront aberration function $W(p,q;\lambda,z_\mathrm{acc})$ to be used in Eq.~\ref{eq:PupilFunc}. 
where $\mathrm{PSF_{polyDL}}(x,y)$ is defined as the special retinal PSF for the eye whose monochromatic aberration coefficients are all zero and is in-focus at the CDP ($z_\mathrm{acc} = z_\mathrm{CDP}$) while the accommodation-related spherical aberration and the wavelength-dependent defocus due to the LCA are present for the sake of consistency in the calculation of polychromatic PSFs. 

\begin{figure}
    \centering
    \includegraphics[width=110mm]{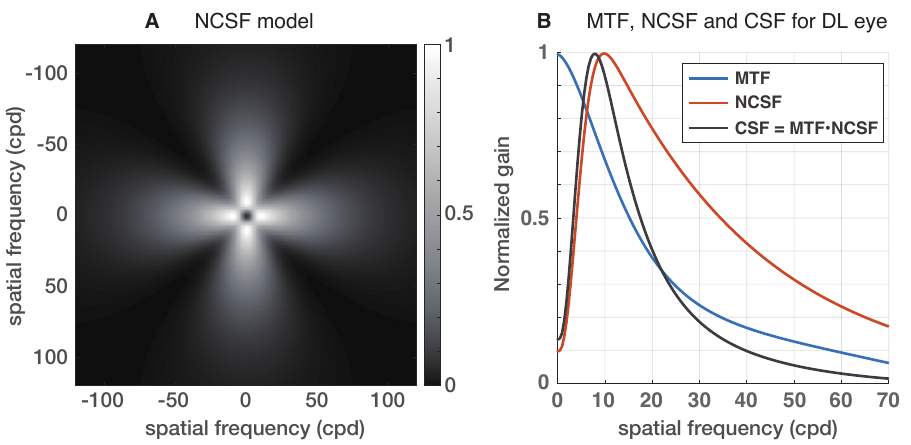}
    \caption{\textbf{A)} The NCSF model by Watson \& Ahumada (2005, 2012). The model is a two-dimensional real function that describes the neural portion of the contrast sensitivity function (CSF) as a filter. \textbf{B)} A cross section of the NCSF model, MTF, and CSF at vertical frequency = 0. Only half of one row is shown due to symmetry. The MTF (blue line) is calculated for the diffraction-limited eye with a 3-mm pupil. The CSF (black line) is a product of the MTF and the NCSF.}
    \label{fig:CSF}
\end{figure}

The NCSF is conceptually the neuro-retinal portion of the contrast sensitivity function (CSF), assuming that the CSF is the product of the eye's MTF and the NCSF as shown in Fig.~\ref{fig:CSF}B~\cite{Michael2011-df}. The NCSF is obtained either by measuring the contrast sensitivity to interference fringes that are directly formed on the retina~\cite{Campbell1965-ii} or by dividing the CSF by the MTF~\cite{Michael2011-df}. 
%Since we were not able to generate individual NCSFs or CSFs while individual OTFs were obtained from individual wavefront aberrations that followed the statistical population, we used a standard model derived by Watson and Ahumada~\cite{Watson2012-yp,Watson2005-jz}. 
We used the NCSF model derived by Watson and Ahumada as shown in Fig.~\ref{fig:CSF}A~\cite{Watson2012-yp,Watson2005-jz}. 
The NCSF model incorporates the oblique effect, which is the phenomenon that a typical human observer is less sensitive to the oblique gratings than to the horizontal or vertical gratings. 

% To calculate the VSOTF, a neural transfer function (NTF) is required. We used the NTF model derived by Watson and Ahumada~\cite{Watson2012-yp,Watson2005-jz}. The NTF model includes the oblique effect, which is the phenomenon that a typical human observer shows less sensitivity to the oblique gratings than to the horizontal or vertical gratings. 

% \clearpage

\section{Simulation results}

%ES: 1) How about the perceivable spatial resolution for different number of views? We need to discuss the spatio-angular trade-off; we can include (a) figure(s) illustrating such trade-off wrt. number of views (and other possible factors)?
% 2) The effects of display induced "distortions", i.e., diffraction ignoring other aberrations?
% 3) Do we plan to include results for horizontal parallax-only case, and the effects of cross-talk (overlapping and non-overlapping sub-apertures)?
% 4) Retinal defocus blur: do we plan to discuss/present how (accurately) the retinal defocus blur is created for different cases, i.e., num. of views; or do we consider only accommodation (in case it is to be included, we also need to discuss it in the main section)? E.g., considering the scene depth range of 0-4D, CDP of 2D, and image depth of 4D; what happens with the object at 0D when the eye focuses at 4D? Such information should be already available by the through-focus analysis?
%ES_v2
%Ideally, the axes in the figures and other information should have the same font size as the text; if this is not feasible, they should be readable at least
%Fig.4: the space is not well utilized, why the lines don't continue to larger diopters?
%Fig.5: indicate the amount shift on each figure
%Let's also put results for example 2D (stimuli) images.
%A subsection for the metric itself, e.g.: VSOTF vs visual strehly ratio: address the discrepancy

In this section, (1) predictions of accommodation response to LF images and (2) the optical resolution of the images are reported for a wide range of depths rendered by an integral-imaging-based LF display with various viewpoint densities. 
Through-focus analysis was employed to predict the accommodation response, and optical resolutions on the retina at various conditions were compared between each of them at the best focused state found by the through-focus analysis. 

The CDP depth of the LF display model was fixed at 2 dioptres. The three viewpoint densities, i.e. 2$\times$2, 3$\times$3, and 4$\times$4 viewpoints within the eye pupil of a 3-mm diameter, were tested. The rendered image depth was one of nine depths in front of and behind the CDP, namely $-2.0$D, $-1.5$D, $\dots\text{,}$ $0$D, $\dots\text{,}$ or $+2.0$D relative to the CDP depth. 
The intervals of the nominal accommodation distances for the through-focus analysis were set to be $0.2$D. 
The spectrum of visible light was sampled at 400 nm and 10 nm increments up to 700 nm. We set the reference wavelength at 550 nm, at which the luminosity function approximately reaches its peak~\cite{Ravikumar2008-ni}. 
The whole simulation was implemented by customising the ISETbio toolbox~\cite{Cottaris2019-dk} on MATLAB. 

\begin{figure}
    \centering
    \includegraphics[width=90mm]{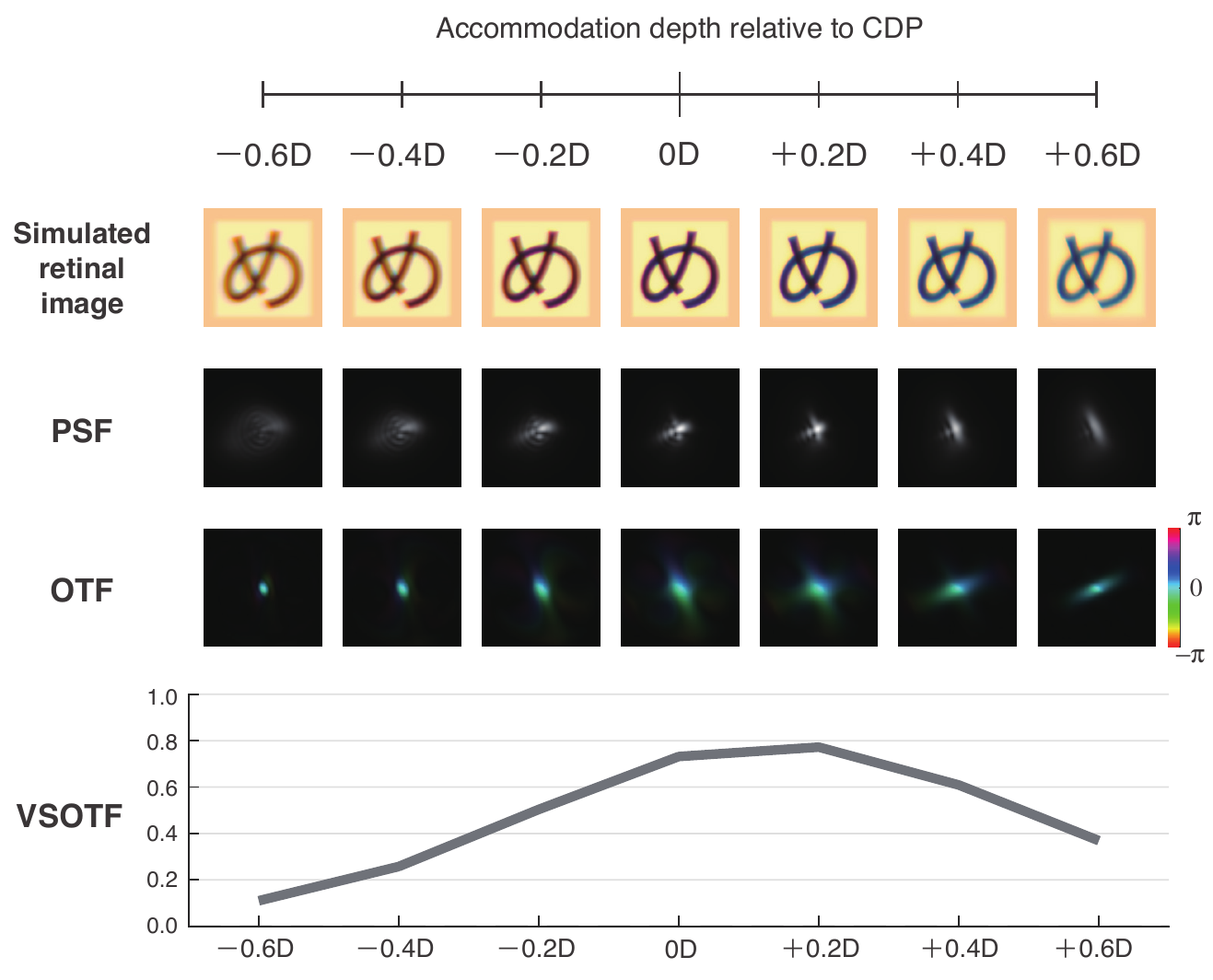}
    \caption{Example of through-focus analysis for one open pupil of an aberrated eye. The CDP depth was 2D. OTFs are visualised in a manner that the brightness and hue correspondingly depict the OTF's modulus and angle. }
    \label{fig:exampleResultsEye1}
\end{figure}

\subsection{Prediction of accommodation response} 

%Accommodation responses to rendered images were predicted by the through-focus analysis of VSOTF for several view densities and the wide range of the image depth. 
Fig.~\ref{fig:exampleResultsEye1} shows an example of the through-focus analyses we conducted. In this case, an analysis for one open pupil of an aberrated eye is shown, thus the rendered image depth was only the CDP depth. PSFs were first computed for several accommodation states that had varying nominal accommodation distance $z_\mathrm{acc}$ in front of and behind the image depth and the CDP depth. The dioptric interval of the nominal accommodation distances was 0.2D. The frequency domain of PSFs, namely OTFs, were then computed and finally VSOTF values were obtained for each OTF. The nominal accommodation distance at which the VSOTF takes the peak value is the predicted accommodation response. Therefore, in Fig.~\ref{fig:exampleResultsEye1}, an accommodation response is predicted at $+0.2$D relative to the image depth. 
This analysis was repeated for all combinations of the eyes, viewpoint densities, and rendered image depths. 

\subsubsection{Natural view}

\begin{figure}
	\centering
	\includegraphics[width=118mm]{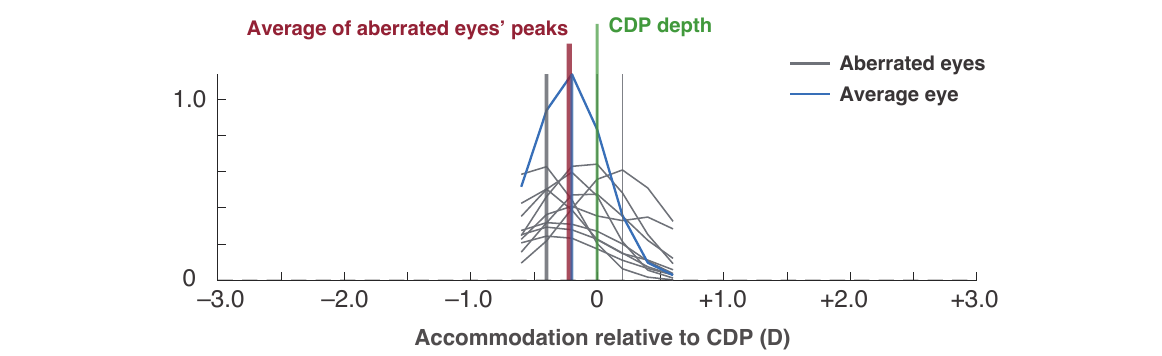}
	\caption{Through-focus VSOTF for natural view. The grey curves indicate metric values as functions of relative accommodation, namely additional defocus to the rendered image depth. The grey vertical lines show their peaks. The red vertical line is the average of the peak positions over the eyes. The blue curve and vertical line represent the metric values and its peak for the average eye. The image depth was that of the CDP (2 dioptres).}
	\label{fig:TFopenpupil}
\end{figure}

Firstly, for reference, we conducted the simulation for the case of natural view, where no subaperture was set and thus the `rendered' image depth was only the CDP depth. 
Fig.~\ref{fig:TFopenpupil} shows the through-focus VSOTFs calculated for the aberrated eyes and the average eye; their peak positions are shown as vertical lines, which indicates the predicted accommodation depths for the eyes. 
The aberrated eyes had generally different through-focus response profiles from each other (the grey curves), but none of the ten eyes was predicted to focus at a depth apart from the image depth ($=$ the CDP depth) by more than $0.2$D. 
To see the average response of the aberrated eyes, the average position of the peaks for the aberrated eyes is shown with the vertical red line. 
The average response was close to $0$D with a slight negative shift in dioptres. It means the eyes were, on average, predicted to focus at a depth close to the CDP when observing a point on the CDP naturally. 

The through-focus VSOTF for the average eye (the blue curve) was much greater than that for the aberrated eyes around its peak and almost reaching the value of 1, suggesting an eye with the average aberration is very close to the diffraction-limited eye when it is in focus. 
The predicted accommodation response for the average eye was, however, essentially the same as the average response of the aberrated eyes. 

\subsubsection{Rendered 3D image} 

Next, we simulated retinal PSFs of images rendered at a wide range of depths relative to the CDP, and accommodation responses to the rendered image were predicted by the through-focus analysis of VSOTF. 
The simulation was conducted for three configurations where 2$\times$2, 3$\times$3, or 4$\times$4 viewpoints were in the pupil of 3 mm diameter. 

\begin{figure}
    \centering
    \includegraphics[width=132mm]{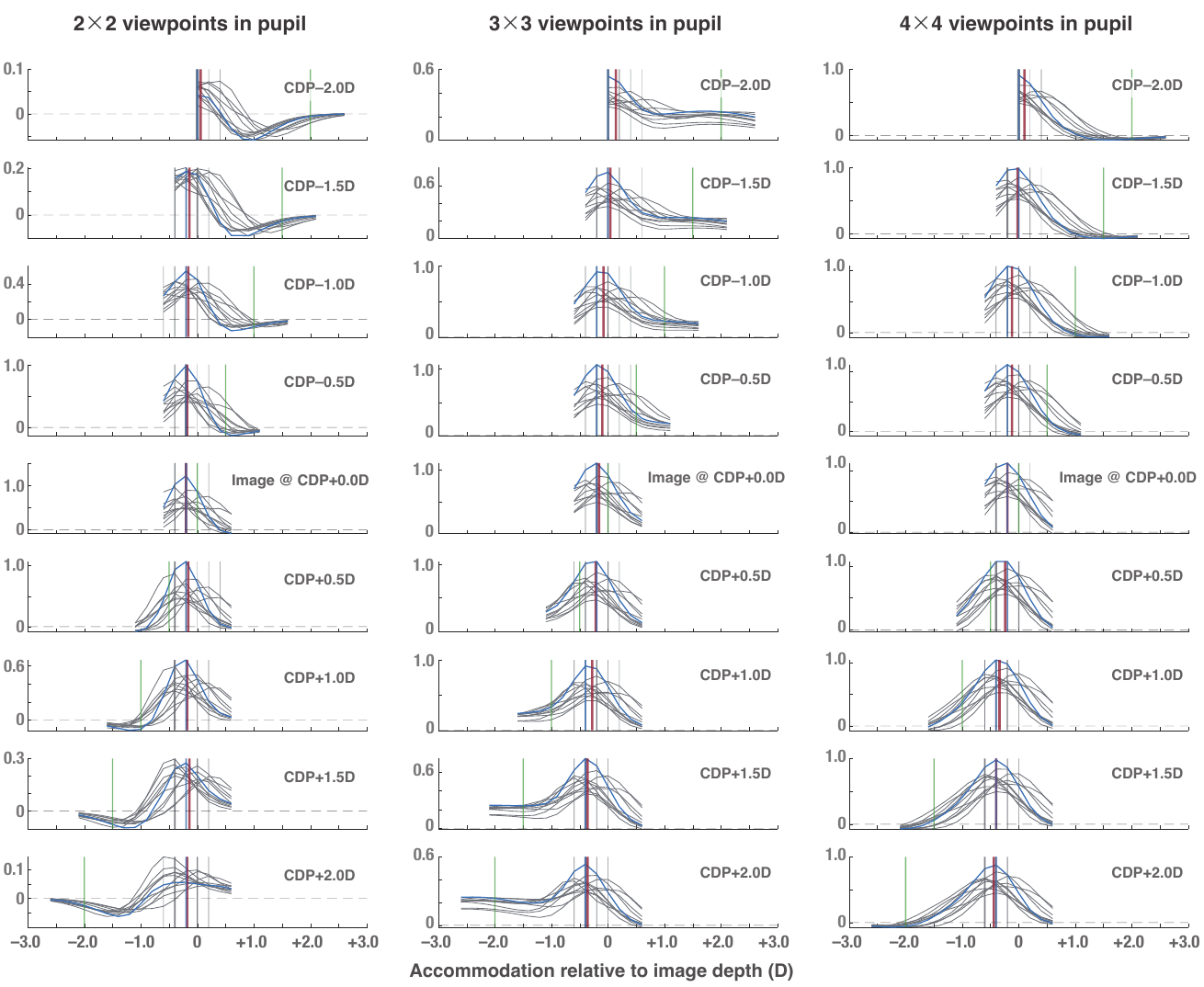}
    \caption{Through-focus VSOTF for the viewpoint densities of 2$\times$2, 3$\times$3, and 4$\times$4 viewpoints in a 3-mm pupil. Data are plotted in the same manner as Fig. \ref{fig:TFopenpupil}. The green vertical lines indicate the CDP depth. Note that some parts of the through-focus ranges are missing in some panels on the top because we assumed the eyes would not accommodate hyperopically, namely its absolute accommodation distance should not be `farther than infinity'.} 
    \label{fig:TFviewpoints}
\end{figure}

\begin{figure}
    \centering
    \includegraphics[width=132mm]{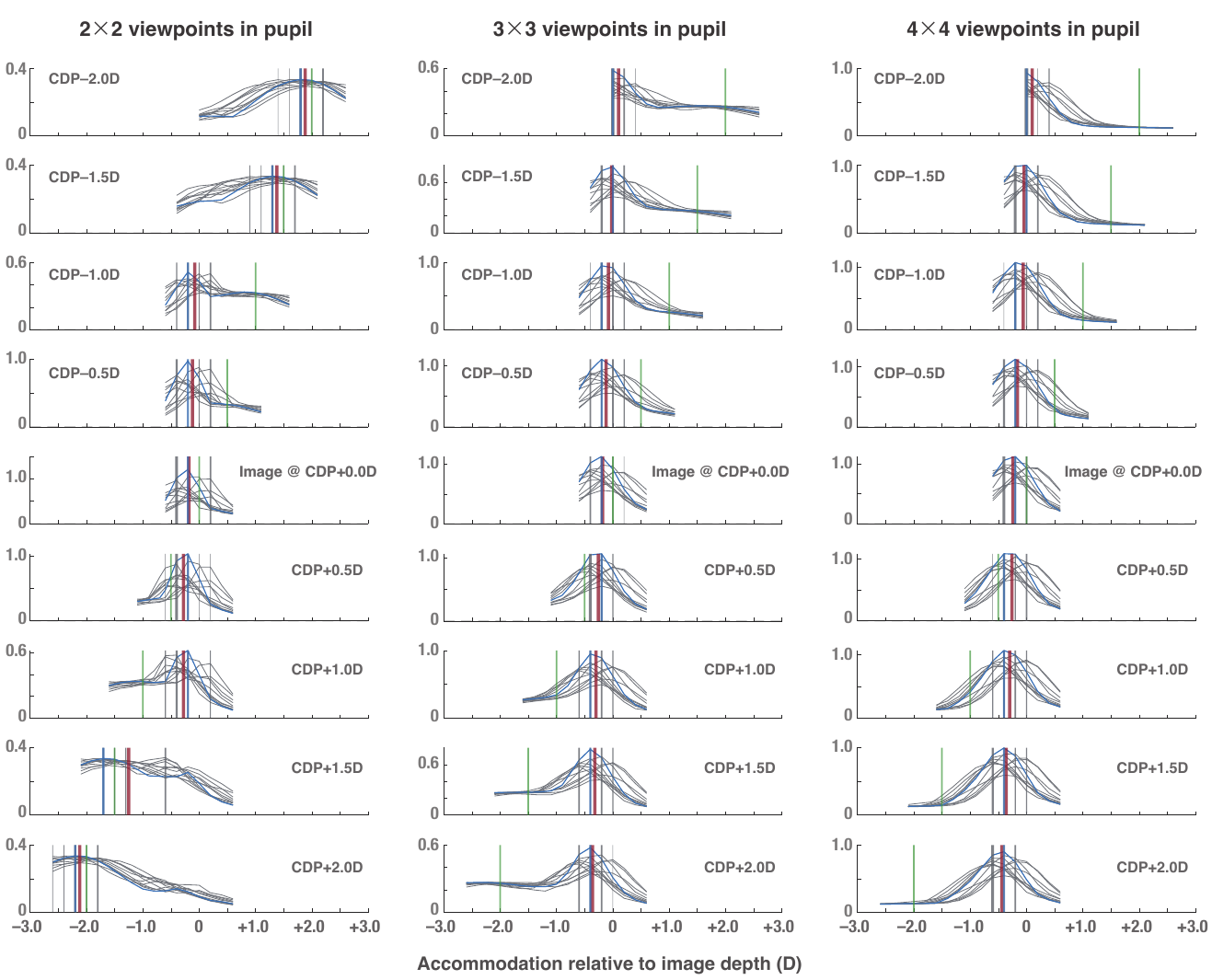}
    \caption{Through-focus Strehl ratio for viewpoint densities of $2{\times}2$, $3{\times}3$, and $4{\times}4$ viewpoints in a 3-mm pupil. Data are shown in the same manner as Fig.~\ref{fig:TFviewpoints}.}
    \label{fig:TFSR}
\end{figure}

Fig.~\ref{fig:TFviewpoints} shows the through-focus plots of VSOTF for each image depth and view density. 
The mean predicted responses of the aberrated eyes were close to the rendered image depth for all viewpoint densities and image depths. 
In other words, the accommodation error, i.e. the dioptric difference between the image depth and the predicted accommodation distance, was predicted to be small regardless of the rendered image depth and the viewpoint density. 
The accommodation errors tended to be small but non-zero and constantly negative. This means that the aberrated eyes were predicted to focus at depths slightly farther than the image depth regardless of the rendered depth of the image. 

The tendency of the negative errors may be due to the effect of the spherical aberration in the eye. That is to say, focusing at slightly farther than the nominally in-focus depth may improve the retinal image quality, or increase VSOTF in this case, under presence of the spherical aberration~\cite{Lopez-Gil2010-hf,Lopez-Gil2013-ap,Thibos2013-mf}. 
The constant accommodation errors across the image depths contradict the previously reported simulation results in which the accommodation error was reported to increase as the depth difference between the rendered image and the CDP grows. 
%ES_v2: which particular works? Put ref. Any possible effects of finite pixel size in those work, if not, what are the possible reasons (maybe in the discussion?)

The accommodation responses predicted for the average eye (the blue vertical lines in Fig.~\ref{fig:TFviewpoints}) were very close to the average of the predicted accommodation responses for the aberrated eyes (the red vertical lines in the same figure). 
This similarity suggests that a through-focus analysis of VSOTF on the average eye may be sufficient to predict accommodation responses, although the average eye is very close to a diffraction-limited system and thus poorly represents aberrated real eyes. 

In summary, the accommodation responses were predicted to be close to the rendered image depths, and the predicted accommodation errors were roughly constant across the image depths. In addition to that, increasing the viewpoint density did not affect the accommodation errors, at least in the configurations we tested. 
It can straightforwardly be interpreted that even the lowest viewpoint density we tested, namely 2$\times$2 viewpoints in the 3-mm diameter pupil, is fairly enough to elicit an accommodation response as correctly as in the natural view.  

These observations generally contradict the findings that have previously been reported~\cite{Huang2017-bc,Xu2020-rc,Qin2020-bi,Qin2020-db}. 
Huang and Hua predicted smaller accommodation errors in configurations with higher viewpoint densities~\cite{Huang2017-bc}, and Qin et al. reported that systematic accommodation errors towards the CDP depth were predicted from through-focus analyses of the Strehl ratio, even though a large number of viewpoints (45 viewpoints over a 4-mm pupil) was assumed in their simulation~\cite{Qin2020-bi,Qin2020-db}. 
In the current study, however, the predicted accommodation responses did not show any clear difference between the viewpoint densities, and the predicted accommodation errors were small and constant across the rendered image depths. 

There are two possible points that may explain the difference. 
One is the effect of display diffraction and aberration, which were ignored in the simulation in the current study but included in the simulations in all previous studies. 
Including them into the model must always make the elemental PSFs more blurred, since a point source at the CDP is an idealised simplification of the image of a point source on the rendering plane. 
Nevertheless, it does not directly imply that including the effect of display diffraction and aberration into the simulation makes the results closer to these in the previous studies. 
%ES_v2:what do you mean with above sent.?
The other is the prediction methods of accommodation responses. 
In the current study, VSOTF was used as the predictor metric for accommodation response, while the previous studies used through-focus Strehl ratios and values picked from through-focus MTFs. 

We also analysed through-focus Strehl ratios for the viewpoint densities and rendered image depths (Fig.~\ref{fig:TFSR}). 
The accommodation responses predicted from through-focus Strehl ratios were essentially identical to those from the through-focus VSOTF except the cases for the viewpoint density of 2${\times}$2 viewpoints in the pupil. 
That is to say, the through-focus Strehl ratio predicted that the accommodation responses would be close to the image depths rather than the CDP depth in the cases for the viewpoint densities of 3$\times$3 or 4$\times$4 viewpoints in a 3-mm pupil. 
For the viewpoint density of 2$\times$2 viewpoints in the pupil, however, the results were largely different from the other viewpoint densities. Specifically, the predicted accommodation depths were close to the image depths only if the rendered images were in a depth range of ${\pm}1.0$D relative to the CDP; otherwise, the predicted accommodation responses were close to the CDP rather than the image depths. 
The tendency of small negative accommodation errors was observed in the through-focus Strehl ratio as well as in the through-focus VSOTF except in the cases where the image depths were apart from the CDP by more than 1.0D for the viewpoint densities of 2$\times$2 viewpoints in the pupil. 
The predicted accommodation responses on the average eye were very similar to those on the aberrated eyes, as observed also in the through-focus analysis of VSOTF. 

\begin{figure}
    \centering
    \includegraphics[width=111mm]{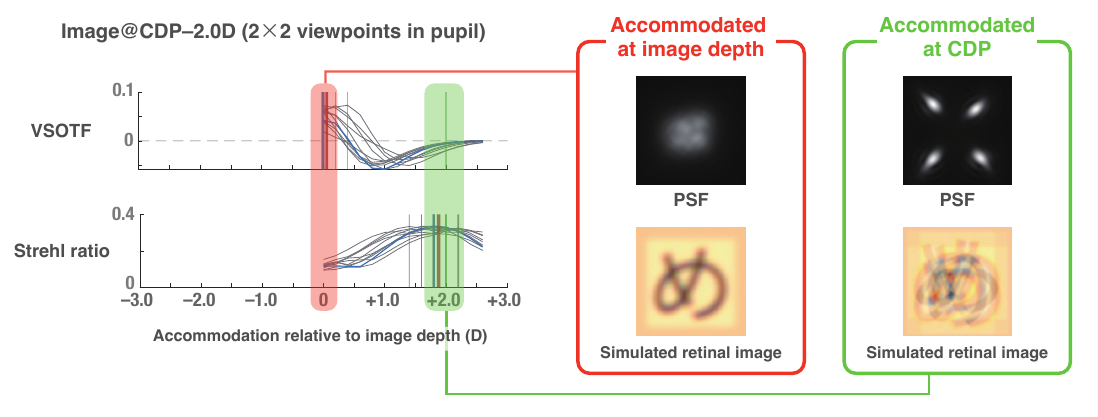}
    \caption{Through-focus VSOTF and Strehl ratio for an image rendered at the CDP depth $-2.0$D and PSFs and simulated retinal images at the two accommodation states in the through-focus range where the eye accommodates at the rendered image and the CDP. The PSFs and retinal images are for one of the aberrated eyes. }
    \label{fig:SRvsVSOTF}
\end{figure}

Fig.~\ref{fig:SRvsVSOTF} shows an example case in which the predictions from the through-focus VSOTF and Strehl ratio were largely different; VSOTF predicted the accommodation response around the image depth, while the Strehl ratio predicted the response at the CDP depth. 
When the eye accommodated at the image depth, the PSF spread widely and thus the simulated retinal image was blurred. On the other hand, when the CDP was accommodated, the PSF consisted of four separate and clear peaks, hence the retinal image appeared as a superposition of four clear images with positional differences, which eventually looked hardly recognisable. 
Since only the peak amplitude of a PSF defines the Strehl ratio, its highest value seemed to be obtained in the case when the CDP was accommodated. 
There is much evidence to support the higher validity of VSOTF over the Strehl ratio as a predictor of accommodation response, but all of these studies, to the best of our knowledge, tested it for eyes with a natural pupil. 
Hence, the validity of VSOTF is not perfectly supported in predicting the accommodation response from `irregular' PSFs that are rendered by an LF display. 
However, this also applies to other metrics such as the Strehl ratio. Considering that the Strehl ratio is usually used only for optical systems with little aberration, the Strehl ratio may not be a suitable metric to predict the accommodation response from PSFs rendered by an LF display. 

%\clearpage
\subsection{Optical resolution of rendered image}

\begin{figure}
    \centering
     \includegraphics[width=133mm]{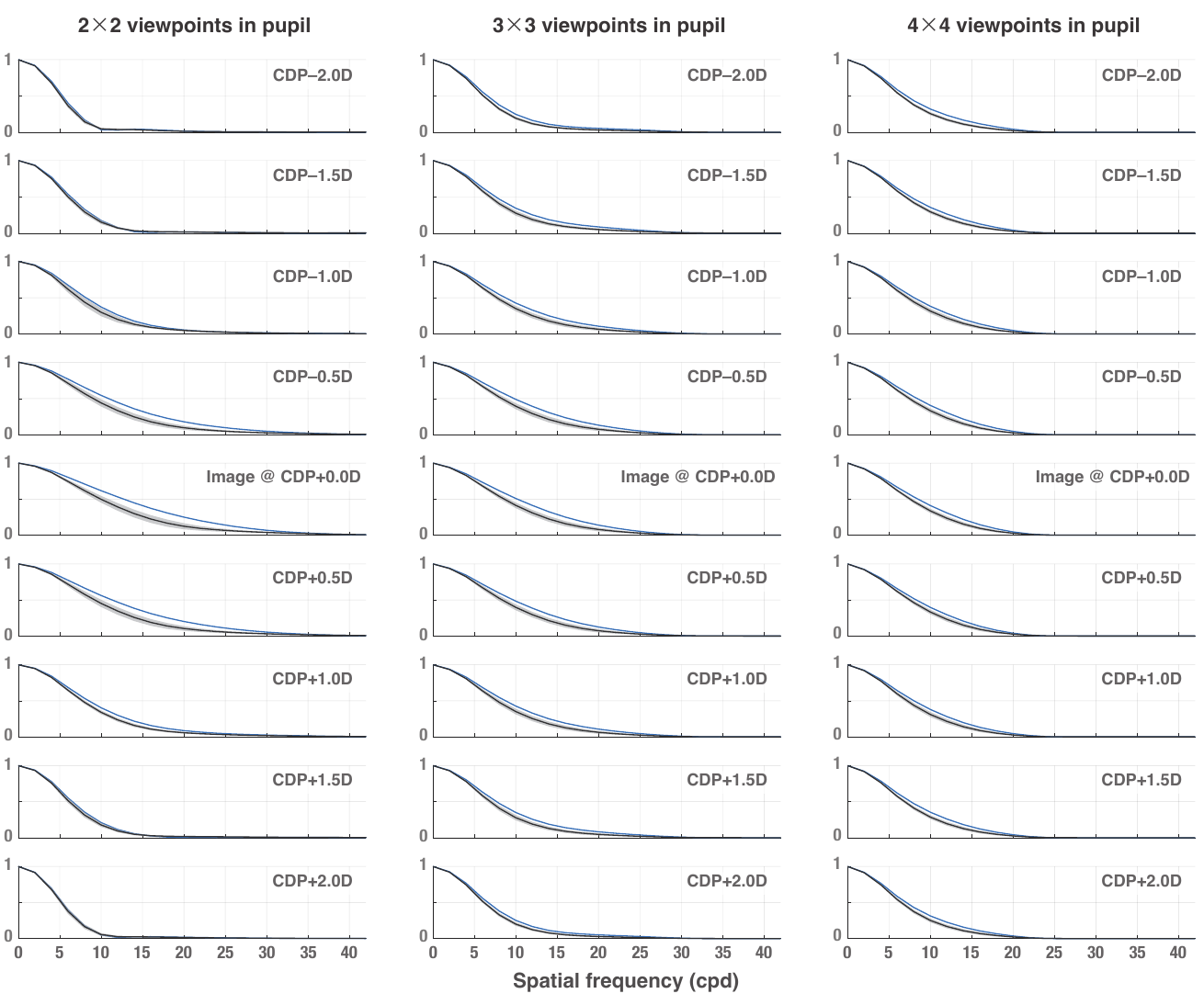}
    \caption{Radial MTFs of the retinal images at predicted accommodation by VSOTF for each image depth and viewpoint density. The black plots with grey shades are the mean and $\pm{}1$ standard deviation of the radial MTFs for the aberrated eyes. The blue plots are the radial MTFs for the average eye. }
    \label{fig:resolution}
\end{figure}

%The optical resolution of the rendered images on retina was analyzed across the image depths and viewpoint densities. 
%To assess the optical resolution, the in-focus retinal PSF that gave the best VSOTF value was picked from the through-focus PSFs at each image depth and viewpoint density for each eye. Radial MTFs were then calculated from the in-focus retinal PSFs, which describe how much frequency components in the image are filtered~\cite{Thibos2002-jr,Watson2013-yl}. 
The two-dimensional MTFs for the in-focus accommodation states predicted by the VSOTF were averaged across meridians, providing \emph{radial MTFs}~\cite{Thibos2002-jr,Watson2013-yl}, so that the MTFs were visualised and compared with each other more simply. 
Fig.~\ref{fig:resolution} shows the obtained radial MTFs, which describe the optical resolution on the retina for each rendered image depth and viewpoint density. The mean MTF for the aberrated eyes and the MTF for the average eye are separately plotted. 
The MTFs for the average eye were better than the mean MTFs for the aberrated eyes in almost all cases. This indicates that the simulation on the eye model with the average aberration overestimates retinal optical resolution. 

%ES_v2: highlight/denote, e.g, 0.1 MTF: resolution?

In general, better MTFs were simulated for the cases with the smaller depth difference between the rendered image and the CDP, especially when the viewpoint density was low. This tendency was less clear for the cases with higher viewpoint densities. 
Specifically, for the cases with the lowest viewpoint density --- namely 2$\times$2 viewpoints in the pupil --- the optical resolution drastically dropped as the depth difference between the rendered image and the CDP grew, even though the retinal image was focused. 
It can be interpreted that the depth of field (DOF) of the LF display, or the range of a rendered depth where the resolution of the in-focus retinal image is kept high, is shallow for a configuration with a low viewpoint density. 
On the other hand, with the high viewpoint density, the drop of the in-focus optical resolution related to the rendered image depth relative to the CDP was less distinctive. In other words, a wider DOF was simulated for a configuration with a high viewpoint density. 

\begin{figure}
    \centering
    \includegraphics[width=117mm]{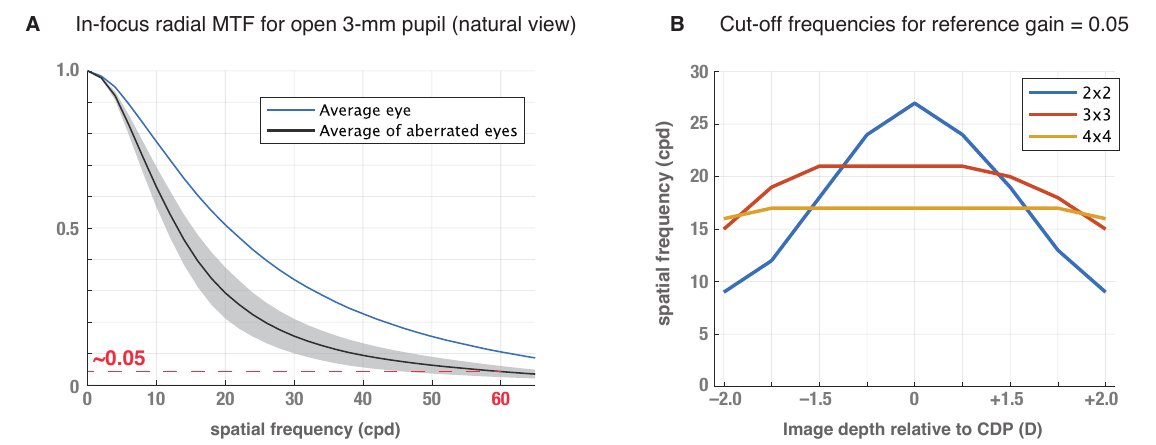}
    \caption{\textbf{(A)} Radial MTF for open 3-mm pupil at the in-focus state predicted by VSOTF. The blue line represents the radial MTF for the average eye. The black line and grey shaded area show the mean and $\pm$1 standard deviation of the radial MTFs for the aberrated eyes. \textbf{(B)} Cut-off frequencies for reference gain of 0.05 for the three viewpoint densities.}
    \label{fig:cofreq}
\end{figure}

To visualise the relation between the viewpoint density and the optical resolution on the retina, we set the reference contrast gain of 0.05 as a threshold and plotted the cut-off frequencies for the three viewpoint densities (Fig.~\ref{fig:cofreq}). 
We chose the reference gain of 0.05 because the contrast gain dropped to around that value on average in the aberrated eyes at 60 CPD, which is dogmatically the highest spatial frequency resolvable for human observers (Fig.~\ref{fig:cofreq}A). 
Fig.~\ref{fig:cofreq}B shows the cut-off frequencies for the three viewpoint densities as functions of the image depth relative to the CDP. 
As already described, a configuration with the low viewpoint density (2$\times$2 viewpoints) was predicted to achieve a high resolution, but the DOF was small. On the other hand, a configuration with a high viewpoint density (4$\times$4 viewpoints) can achieve low resolution at best, but the DOF was greater than the lower viewpoint densities. 

The best optical resolution, which was achieved when the image was rendered at the CDP depth, was better in the cases with the low viewpoint density. A drop of the maximum resolution with increasing viewpoint density was also reported previously~\cite{Huang2017-bc}, but the drop observed in the current study was not as drastic as in their report. 

\subsection{Summary}

We simulated through-focus PSFs for a wide range of the image depths and the viewpoint densities of 2$\times$2, 3$\times$3, or 4$\times$4 viewpoints in a 3-mm diameter pupil assuming no effect of display diffraction and aberration. 
The accommodation responses to the rendered images were predicted by the through-focus analysis of VSOTF, and the optical resolution on the retina was also assessed. 

The predicted accommodation responses were close to the rendered image depths even for the cases where the image depth was apart from the CDP depth. Importantly, the predicted accommodation errors were constantly small across the viewpoint densities. This means that even the lowest viewpoint density we tested --- namely 2$\times$2 viewpoints in the pupil --- may be enough to elicit correct accommodation to rendered images at a wide range of depths if the LF display had the ideal optics so that the effects of its diffraction and aberration were negligible. 

The analysis of the optical resolution on the retina showed the advantage of a wide DOF for the LF display with a high viewpoint density. In other words, increasing the viewpoint density extended the range of a rendered image depth in which the best retinal image can be kept. 
% Furthermore, increasing viewpoint density did not seem to impair the maximum resolution of the retinal image severely. 

To sum up, under the assumption of the absence or negligible effects of display diffraction and aberration, it was suggested that the viewpoint density may not be needed to be that high to elicit the correct accommodation responses to rendered images; however, it was also inferred that a low viewpoint density may limit the range of a rendered depth in which the in-focus resolution is kept high on the retina. 
%ES_v2 "that high"? "low viewpoint density"?: these are relative, do not really tell concrete information/observation. 

\section{Limitations of the study}

The current study is limited principally at several points. 
The most important limitation is the absence of display diffraction and aberration. 
As discussed earlier, ignoring these effects and assuming an ideal point source at the CDP must overestimate the optical resolution on the retina. 
Hence, the simulation results estimate the retinal image resolution only for an LF display with little effects of diffraction and aberration, which seems to be hardly achievable. 
Simultaneously, it can be inferred that no real LF displays can achieve a retinal image resolution that is better than the results simulated in the proposed way. 
On the other hand, the effects of display diffraction and aberration on accommodation response cannot be simply guessed from the logic that including them would make the elemental PSFs less sharp, because the visual system always reacts to their superposition. 

Another limitation is the absence of pixels in the model. 
In the current study, the point source was assumed to be infinitesimal and positional sampling was ignored. 
However, in a practical LF display, the image resolution is always regulated by the display's pixel sampling. 
Pixel sampling strongly affects the image resolution a viewer may observe in combination with the rendered depth~\cite{Hoshino1998-xo,Zwicker2007-ul}. 
Specifically, the pixel sampling determines the depth range in which the best pixel resolution can be obtained for the viewer. 
A further study is needed to integrate the optical effects of diffraction and aberration in a display and human eye into the discussion of the pixel resolution and the rendered depth, although an initial attempt has been made by Huang and Hua by extending their study~\cite{Huang2019-gd}. 

The last important limitation to be pointed out is the paraxial approximation. 
The calculation of PSFs in the proposed framework gives an accurate simulation only in the paraxial region. 
In other words, the calculation cannot simply be used to assess the optical image on the peripheral retina, which is formed by off-axis light rays. 
In addition to that, because the retina is not spatially uniform physiologically and perceptually, the functional role of the non-foveal or peripheral region in accommodation must be different from that of the foveal region~\cite{Labhishetty2019-lt}. 
Therefore, involving the functional role of the non-foveal or peripheral retina in accommodation may require not only an appropriate calculation method for the retinal image, but also an evaluation method that correctly reflects the characteristics of the peripheral visual field. 
%ES_v2: we are not really discussing the retinal defocus blur in this paper, so non-paraxial approx. or what happens at periphery is not really a limitation. I would remove this paragraph.

\section{Conclusion}

In the current study, we proposed a novel simulation framework to model an LF display and an observer's eye to simulate the retinal image, which is totally free from ray-tracing. 
The proposed framework has the advantages of (1) being free from ray-tracing, thus computationally efficient; (2) having the capability to include realistic aberration patterns of the human eye population; and (3) ensuring the rigorous modelling of chromatic effects in the visual system. 
% flexibility to separate the effects of display diffraction and aberration 

In addition to that, an optical metric that is known to predict accommodation response well was used to assess whether the accommodation would be elicited correctly. 
The simulation based on the proposed model showed that accommodation was expected to be elicited fairly close to the rendered depths even in a configuration with a relatively low viewpoint density, e.g. 2$\times$2 views within a 3 mm pupil. 
However, increasing the viewpoint density seemed to extend the depth range of the rendered image in which in-focus retinal resolution is kept high. %while it did not appear to impair the maximum in-focus resolution. 
%ES_v2: Future work?

\section*{Disclosures}

The authors declare no conflicts of interest. 

\section*{Data Availability}

Data underlying the results presented in this paper are available in Refs.~\cite{Thibos2002-en,Thibos2009-ba}. 

\section*{Funding}

This research project has received funding from the European Union's Horizon 2020 research and innovation programme under Marie Sk\l{}odowska-Curie grant agreement No 764951. 
The project is also partially supported by the Academy of Finland research project “Modeling and Visualization of Perceivable Light Fields”, decision number 325530. 

%%%%%%%%%%%%%%%%%%%%%%% References %%%%%%%%%%%%%%%%%%%%%%%%%

% Add references with BibTeX or manually.
% \cite{Zhang:14,OSA,FORSTER2007,Dean2006,testthesis,Yelin:03,Masajada:13,codeexample}

%%%%%%%%%% If using BibTeX:
\bibliographystyle{osajnl}
\bibliography{myref,YutaCustomBIB}

%%%%%%%%%% If preparing manually:
% \newcommand{\enquote}[1]{``#1''}

% \bibitem{Zhang:14}
% Y.~Zhang, S.~Qiao, L.~Sun, Q.~W. Shi, W.~Huang, L.~Li, and Z.~Yang,
%   \enquote{Photoinduced active terahertz metamaterials with nanostructured
%   vanadium dioxide film deposited by sol-gel method,}
%   {\protect\JournalTitle{Optics Express}} \textbf{22}, 11070--11078 (2014).

% {Optical Society}, \enquote{{OSA Publishing},}
%   \url{http://www.osapublishing.org}.

% \bibitem{FORSTER2007}
% P.~Forster, V.~Ramaswamy, P.~Artaxo, T.~Bernsten, R.~Betts, D.~Fahey,
%   J.~Haywood, J.~Lean, D.~Lowe, G.~Myhre, J.~Nganga, R.~Prinn, G.~Raga,
%   M.~Schulz, and R.~V. Dorland, \enquote{Changes in atmospheric consituents and
%   in radiative forcing,} in \enquote{Climate Change 2007: The Physical Science
%   Basis. Contribution of Working Group 1 to the Fourth assesment report of
%   Intergovernmental Panel on Climate Change,}  S.~Solomon, D.~Qin, M.~Manning,
%   Z.~Chen, M.~Marquis, K.~B. Averyt, M.~Tignor, and H.~L. Miler, eds.
%   (Cambridge University Press, 2007).

% \end{thebibliography}

\end{document}